\documentclass[]{aa}

\usepackage{graphicx}

\newcommand{\ar}[3]  {#1$^{\rm h}$#2$^{\rm m}$#3$^{\rm s}$}
\newcommand{\dec}[3] {#1$^{\circ}$#2$'$#3}
\newcommand{\jr}[2] {\mbox{$J$=#1$\rightarrow$#2}}
\newcommand{\tr}[2] {\mbox{#1$\rightarrow$#2}}

\begin{document}

\title{Multitransitional observations of the CS core of L673}
\author{O.\ Morata
\and J.~M.\ Girart
\and R.\ Estalella
}
\institute{Departament d'Astronomia i Meteorologia, Universitat de Barcelona,
Av.~Dia\-go\-nal, 647, E-08028, Barcelona, Spain}
\date{Received / Accepted}

\offprints{Oscar Morata\\ \email{oscar@am.ub.es}}

\abstract{ 

A multitransitional study with the BIMA interferometric array was carried out
toward the starless core found in the L673 region, in order to study the
small--size structure of the cores detected with previous single--dish
observations, which provides us with a test of the predictions of the chemical
model of Taylor et al. (\cite{tayloru}; \cite{taylord}). We detected emission
in the CS (\jr{2}{1}), N$_2$H$^+$ \mbox{($J$=1$\rightarrow$0)}, and HCO$^+$
\mbox{($J$=1$\rightarrow$0)} lines. Several clumps of size $\la0.08$ pc were
found for each line distributed all over the region where previous single-dish
emission was found (Morata et al.\ \cite{morata}). Each molecular transition
traces differently the clump distribution, although in some cases the detected
clumps are coincident. The distribution of the N$_2$H$^+$ emission and the
single--dish NH$_3$ emission are coincident and compatible with an origin in
the same gas.  The large fraction of missing flux measured for the CS
(2$\rightarrow$1) transition can be explained if the cloud is formed by a
clumpy and heterogeneous medium.  Four positions were selected to derive the
abundance ratios [N$_2$H$^+$/CS] and [HCO$^+$/CS] from the molecular column
density determinations, and to compare them with the values predicted by the
chemical model. The model was able to explain the interferometric
observations, and, in particular, the chemical differentiation of the detected
clumps and the coincidence of the NH$_3$ and N$_2$H$^+$ emissions. The lack of
HCO$^+$ towards the two selected positions that trace the more evolved clumps
cannot be accounted for by the model, but it is possibly due to strong
self-absorption. We propose a classification of the studied clumps according
to the stage of chemical evolution indicated by the molecular abundances.

}

\maketitle

\section{Introduction}

Low-mass star formation takes place in dense cores of molecular clouds
(Beichman et al. \cite{beichman}; Benson and Myers \cite{benson}). The
emission of several molecules, such as CS, NH$_3$, and HCO$^+$, known to be
good tracers of high density molecular gas, has been used to study these dense
cores. However, very soon large discrepancies between the emission of these
molecules were found in some sources (Zhou et al. \cite{zhou}; Myers et
al. \cite{myers}). To clarify the intrinsic differences between the emission
of these molecules and how this emission is related to the actual distribution
of the high density gas, we began a systematic comparison between the emission
of the CS \mbox{($J$=1$\rightarrow$0)} and NH$_3$ \mbox{($J,K$)=(1,1)}
transitions under similar conditions of angular resolution (Pastor et
al. \cite{pastor}; Morata et al. \cite{morata}).  The comparison of the
distribution of the CS and NH$_3$ emission in 14 condensations of 11
star-forming regions confirmed the discrepancies. In particular, we found that
there is a separation $\sim0.2$ pc between the emission peaks of both
molecules; regions traced by CS are larger than those traced by NH$_3$; and CS
lines are $\sim 0.5$~km~s$^{-1}$ wider than those of NH$_3$.

\begin{table*}[t]
\caption{Transitions observed}
\label{bimatransitions}
\begin{center}
\begin{tabular}{llrccc}
\noalign{\medskip}
\hline\noalign{\medskip}
         &            &                          & RMS & Spectral& \\ 
Molecule & Transition & \multicolumn{1}{c}{$\nu$} & per channel & resolution &
Detected?\\
         &            & \multicolumn{1}{c}{(GHz)} & (Jy/beam) & (~km~s$^{-1}$) & \\
\noalign{\medskip}
\hline\noalign{\medskip}
HCO$^+$   & \tr{1}{0}                 & 89.18852  & 0.17 & 0.33 & Yes\\
C$_3$S  & \tr{16}{15}               & 92.48849  & 0.18 & 0.32 & No\\
N$_2$H$^+$    & \tr{1}{0}                 & 93.17402  & 0.16 & 0.32 & Yes\\
C$_2$S  & \tr{8$_7$}{7$_6$}         & 93.87010  & 0.18 & 0.31 & No\\
OCS     & \tr{8}{7}                 & 97.30121  & 0.22 & 0.30 & No\\
CS      & \tr{2}{1}                 & 97.98097  & 0.18 & 0.30 & Yes\\
HNCS    & \tr{9$_{1,8}$}{8$_{1,7}$} & 105.74386 & 0.25 & 0.28 & No\\
C$_2$S  & \tr{9$_8$}{8$_7$}         & 106.34774 & 0.24 & 0.28 & No\\
HC$_3$N & \tr{12}{11}               & 109.17363 & 0.26 & 0.27 & No\\
SO      & \tr{2$_3$}{1$_2$}         & 109.25218 & 0.25 & 0.27 & marginally\\
OCS     & \tr{9}{8}                 & 109.46306 & 0.25 & 0.27 & No\\
\noalign{\medskip}
\hline\noalign{\medskip}
\end{tabular}
\end{center}
\end{table*}

To explain these results, we developed a chemical model (Taylor et al.
\cite{tayloru}) in which high density condensations, or dense cores, are
formed by clumps $\la0.1$ pc in size, which would be unresolved at moderate
angular resolution observations such as the ones made in our study, of
different mass, age, size and density. Most of the clumps disperse before
NH$_3$ abundances build up to significant levels, though these clumps contain
substantial CS, so its emission should be observable. A few clumps, those
sufficiently long lived, or being in a more advanced stage of physical and
chemical evolution because of being denser or more massive, form a significant
content of NH$_3$, while CS abundance decreases with time. These clumps would
possibly continue their evolution to eventually form stars. This model would
account for the difference in size and separation between emission peaks of
the CS and NH$_3$ molecules.

Our chemical model explains the differences as a result of the speed at which
the molecules form. Therefore, a classification could be made between {\it
``early-time''} molecules and {\it ``late-time''} molecules, according to the
time at which these molecules reached their peak fractional abundance. We
examined whether there were other potentially observable molecules that should
show extended emission like CS, or more compact emission as NH$_3$ (Taylor et
al.  \cite{taylord}). Several candidates were found for both groups: e.g. HCN,
H$_2$CO or HC$_3$N are found in the CS family, whilst HCO$^+$, SO, NO or
N$_2$H$^+$ are in the NH$_3$ family.

In order to study the small--size structure of the cores detected with the
single--dish observations, which also provides us with a test of the
predictions of the chemical model, we carried out high angular resolution
observations towards L673 of several transitions corresponding to molecules of
both families of species. The L673 high density condensation was previously
detected in the CS \mbox{($J$=1$\rightarrow$0)} transition at moderate angular
resolution by Morata et al. (1997). The L673 CS condensation is located at
$\alpha (J2000)=19^{\rm h}20^{\rm m}52\fs1$, $\delta
(J2000)=+11\degr15'29\farcs5$, at $\sim9\farcm5$ to the southeast of IRAS
19180$+$1116. This condensation is well suited to our purposes for several
reasons. A high density condensation has been detected nearby in the NH$_3$
\mbox{($J,K$)=(1,1)} transition by Sep\'ulveda et al. (\cite{sepulveda}),
which shows the same general behavior found in the other sources of our
survey: a separation from the CS emission peak $\sim2'$, and a NH$_3$
condensation size clearly smaller than the CS condensation size. It is located
nearby, the estimated distance to the L673 cloud is $\sim300$ pc (Herbig and
Jones, \cite{herbig}), which helps in the study of the smaller size structure
of the cloud. Finally, the condensations detected in CS and NH$_3$ show no
signs of tracers of star formation, such as infrared or radio continuum
sources, Herbig-Haro objects or molecular outflows, which indicates that it is
probably a quiescent core, maybe in the first stages of collapse, before
forming a stellar core or a Class 0 star, and thus fulfilling the conditions
of the chemical model we developed.
\vfill
\section{Observations}

The observations of the L673 region were carried out in 1998 May with the
10-antenna BIMA array\footnote{The BIMA array is operated by the
Berkeley-Illinois-Maryland Association with support from the National Science
Foundation.} at the Hat Creek Radio Observatory in the C configuration.  The
phase calibrators were 1751+096 and 3C395.  In order to include the positions
of the CS \mbox{($J$=1$\rightarrow$0)} and NH$_3$ \mbox{($J,K$)=(1,1)}
transitions single-dish emission peaks of the maps by Morata et al. (1997) and
Sep\'ulveda et al. (\cite{sepulveda}), we made a two-point mosaic with each of
the fields centered approximately at the position of these two peaks.  Thus,
one of the fields was centered at the position $\alpha (2000)=19^{\rm
h}20^{\rm m}52{\rm \fs}2$, and $\delta (2000)=+11\degr13'57''$, and the other
located $90''$ to the North.  Three frequency setups were used, centered at
91, 96 and 107~GHz.  The digital correlator was configured to observe
simultaneously several molecular line transitions at moderately high spectral
resolution, $\sim 0.3$~~km~s$^{-1}$.  The target molecular lines were HCO$^+$
\mbox{($J$=1$\rightarrow$0)}, N$_2$H$^+$ \mbox{($J$=1$\rightarrow$0)}, CS
\mbox{($J$=2$\rightarrow$1)}, SO ($J_K=2_3\rightarrow1_2$) and C$^{18}$O
\mbox{($J$=1$\rightarrow$0)}.  System temperatures for the 91, 96 and 107~GHz
setups were in the 180--500~K, 250--700~K and 500--1000~K range, respectively.
The calibration and data reduction were performed using the MIRIAD software
package (Sault, Teuben \& Wright \cite{sault}).  Mosaic maps were done with
the visibility data weighted by the associated system temperatures and using
natural weighting, and applying the primary beam correction.  The resulting
synthesized beam for the 91, 96 and 107~GHz setups were $12\farcs7 \times
9\farcs6$, $PA=-14\degr$; $13\farcs3\times10\farcs5$, $PA=-17\degr$; and
$13\farcs1 \times 9\farcs3$, $PA=-8\degr$, respectively.  The transitions of
all the observed lines, their velocity resolution and the achieved rms noise
with this velocity resolution are listed in Table~\ref{bimatransitions}. For
the continuum emission we used the data from the 96~GHz correlator setup,
which provided a bandwidth of 800~MHz centered at the frequency of
95.9~GHz. No emission was detected at above 8~mJy~beam$^{-1}$ (at a 3-$\sigma$
level).

\section{Results}

\subsection{Morphology of the molecular emission}

We detected emission in three of the transitions: CS (\jr{2}{1}), N$_2$H$^+$
\mbox{($J$=1$\rightarrow$0)}, and HCO$^+$ \mbox{($J$=1$\rightarrow$0)}. We
detected marginally, at $\sim2\sigma$ level, the SO (\tr{2$_3$}{1$_2$})
transition. The remaining transitions of Table \ref{bimatransitions}, which
where observed mainly because they were found inside the frequency range of
the observations, were not detected. It must be noted that all these
undetected transitions have an upper level energy with temperatures higher
than 15~K, and most of them, OCS is the exception, have also a high dipole
moment. Thus, the temperature and density conditions of the L673 core, and the
low abundances expected for these molecules make them very difficult to
detect.

\begin{figure}[t]
\begin{center}
\includegraphics[width=\hsize]{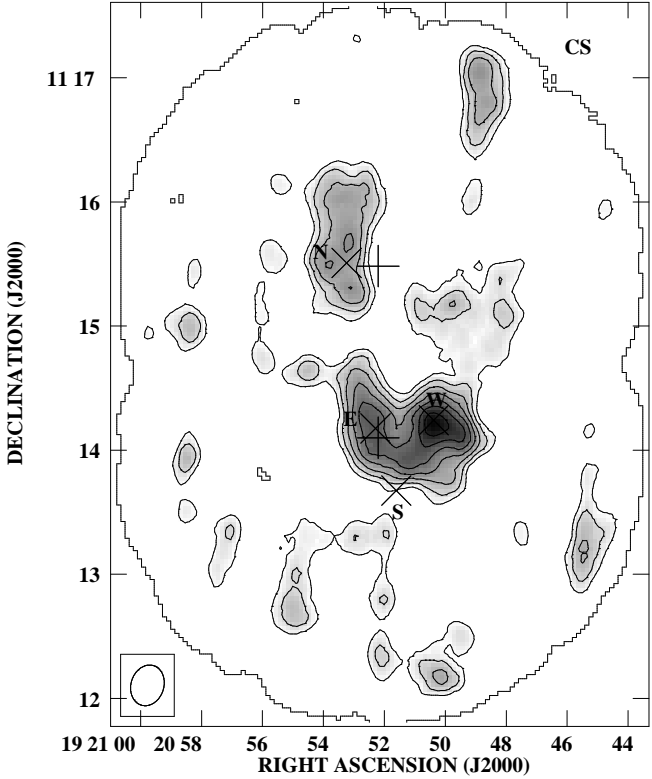}
\end{center}
\caption{Integrated emission of the CS \mbox{($J$=2$\rightarrow$1)} line for
the $V_{\rm LSR}$ range 5.6--9.8 ~km~s$^{-1}$. Contours are 0.96, 1.20, 1.44,
1.68, 1.92, 2.16, 2.40, 2.64 Jy beam$^{-1}$~km~s$^{-1}$. The beam $(20''
\times 15\farcs8)$ is shown in the bottom left-hand corner. The upright
crosses indicate the nominal position of the single-dish emission peak of the
CS \mbox{($J$=1$\rightarrow$0)} line (north) and NH$_3$ \mbox{($J,K$)=(1,1)}
line (south). The tilted crosses indicate the position of the four points
selected for further study (see Table~\ref{tab:positions}). The bordering
contour indicates the two fields observed}
\label{fig:mapcs}
\end{figure}

\begin{figure}[t]
\begin{center}
\includegraphics[width=\hsize]{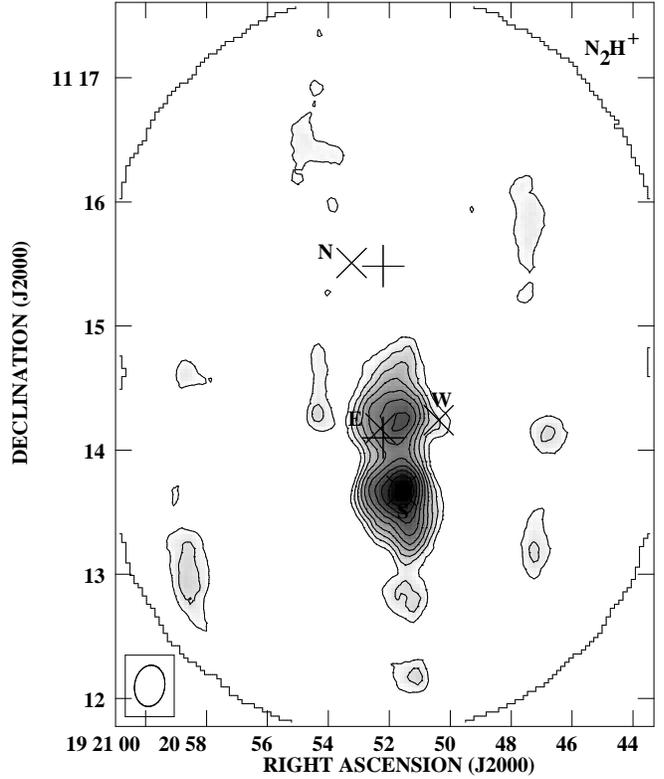}
\end{center}
\caption{Same as in Fig.~\ref{fig:mapcs} for the N$_2$H$^+$
\mbox{($J$=1$\rightarrow$0)} line for the $V_{LSR}$ range $-1.4$--14.2
~km~s$^{-1}$. Contours are 2.55, 3.06, 3.57, 4.08, 4.59, 5.10, 5.61, 6.12,
6.63, 7.14 Jy beam$^{-1}$~km~s$^{-1}$. The beam $(20'' \times 14\farcs6)$ is
shown in the bottom left-hand corner.}
\label{fig:mapn2h}
\end{figure}

\begin{figure}[t]
\begin{center}
\includegraphics[width=\hsize]{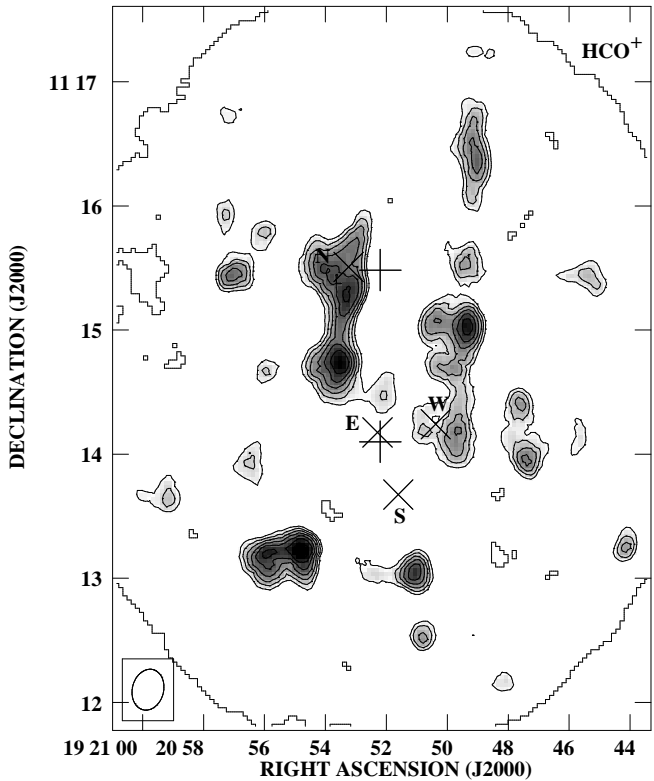}
\end{center}
\caption{Same as in Fig.~\ref{fig:mapcs} for the HCO$^+$
\mbox{($J$=1$\rightarrow$0)} line for the $V_{LSR}$ range 5.6--9.8
~km~s$^{-1}$. Contours are 0.96, 1.08, 1.20, 1.32, 1.44, 1.56, 1.68 Jy
beam$^{-1}$~km~s$^{-1}$. The beam $(20'' \times 15\farcs1)$ is shown in the
bottom left-hand corner. }
\label{fig:maphco}
\end{figure}

Figures~\ref{fig:mapcs} to \ref{fig:maphco} show the maps of the zero-order
moment (integrated intensity) of the emission of the detected molecules. These
maps were obtained after convolving the original maps with a Gaussian
function, with a resulting beam of $\sim20''\times 15''$, in order to make a
more meaningful comparison between the emission of each molecule.

A clumpy distribution of the emission is shown in the integrated intensity map
of the CS (\jr{2}{1}) transition (Fig.~\ref{fig:mapcs}). The more intense
emission is found in two clumps near the position of the single-dish NH$_3$
\mbox{($J,K$)=(1,1)} emission peak: one almost coinciding with its nominal
position, and the other peaking $\sim30''$ to the west. Weaker emission is
found in a clump near the position of the single-dish CS
\mbox{($J$=1$\rightarrow$0)} emission maximum, $\sim1\farcm5$ to the north.

Figure~\ref{fig:mapn2h} shows the integrated intensity map of N$_2$H$^+$
\mbox{($J$=1$\rightarrow$0)}. Two clumps of emission are found, one almost
coincident with the nominal position of the NH$_3$ \mbox{($J,K$)=(1,1)}
emission peak, and the other, more intense, $\sim 0\farcm6$ to the south. No
emission is detected around the position of the CS
\mbox{($J$=1$\rightarrow$0)} single dish emission peak.

The integrated intensity map of the HCO$^+$ \mbox{($J$=1$\rightarrow$0)}
emission (Fig.~\ref{fig:maphco}) shows several isolated emission enhancements
distributed around the positions of the CS and NH$_3$ single-dish emission
peaks. The most intense emission tends to be around the position of the CS
\mbox{($J$=1$\rightarrow$0)} emission peak or in-between the two peaks, with a
N--S elongation. Another intense enhancement is found $\sim1\farcm1$
south-east of the NH$_3$ emission peak nominal position.

\begin{table*}[t]
\caption{Selected positions}
\label{tab:positions}
\begin{center}
\begin{tabular}{lrrl}
\noalign{\bigskip}\hline\noalign{\smallskip}
& \multicolumn{2}{c}{Position} & map counterpart\\ 
\cline{2-3} & $\alpha$ (J2000) & $\delta$ (J2000) & \\
\noalign{\smallskip}\hline\noalign{\medskip}
South (S) & \ar{19}{20}{51.61} & \dec{11}{13}{40\farcs5} & N$_2$H$^+$ \mbox{($J$=1$\rightarrow$0)} south
peak \\ 
East (E)  & \ar{19}{20}{52.29} & \dec{11}{14}{10\farcs5} & CS (\jr{2}{1}) eastern peak \\
West (W)  & \ar{19}{20}{50.38} & \dec{11}{14}{14\farcs5} & CS (\jr{2}{1}) western peak \\  
North (N) & \ar{19}{20}{53.24} & \dec{11}{15}{30\farcs5} & HCO$^+$ \mbox{($J$=1$\rightarrow$0)}
NE enhancement \\ 
\noalign{\medskip}\hline\noalign{\bigskip}
\end{tabular}
\end{center}
\end{table*}

Comparing the integrated intensity maps of the three molecules, we found that
N$_2$H$^+$ emission was more concentrated than CS and HCO$^+$ emission, which
were found more spread all over the region. CS (\jr{2}{1}) emission in the
southern region coincided with the northern clump of N$_2$H$^+$
emission. Moreover, the northern N$_2$H$^+$ emission peak was found between
the two CS (\jr{2}{1}) local emission enhancements. CS (\jr{2}{1}) and HCO$^+$
emission were found to coincide closely at the northern part of the region,
especially around the CS \mbox{($J$=1$\rightarrow$0)} emission peak. On the
contrary, N$_2$H$^+$ and HCO$^+$ emission coincided only marginally in
general. Finally, we were also able to find an emission enhancement in the
marginally detected SO emission very near the position of the N$_2$H$^+$
northern clump.

Table~\ref{tab:positions} lists the four positions we selected in the mapped
region in order to compare the physical parameters of the gas traced by each
molecule. The selected positions were related to the most prominent higher
resolution clumps found in our observations, and corresponded to local
intensity peaks in the emission of the detected molecules. The four positions
were labeled as S, E, W and N, and roughly correspond to the South peak
position of the N$_2$H$^+$ integrated intensity map, the East CS local
maximum, the West CS maximum, and the North enhancement in the HCO$^+$
emission near the CS \mbox{($J$=1$\rightarrow$0)} emission peak, respectively.

\subsection{Kinematic structure}

\begin{figure}[t]
\begin{center}
\includegraphics[width=\hsize]{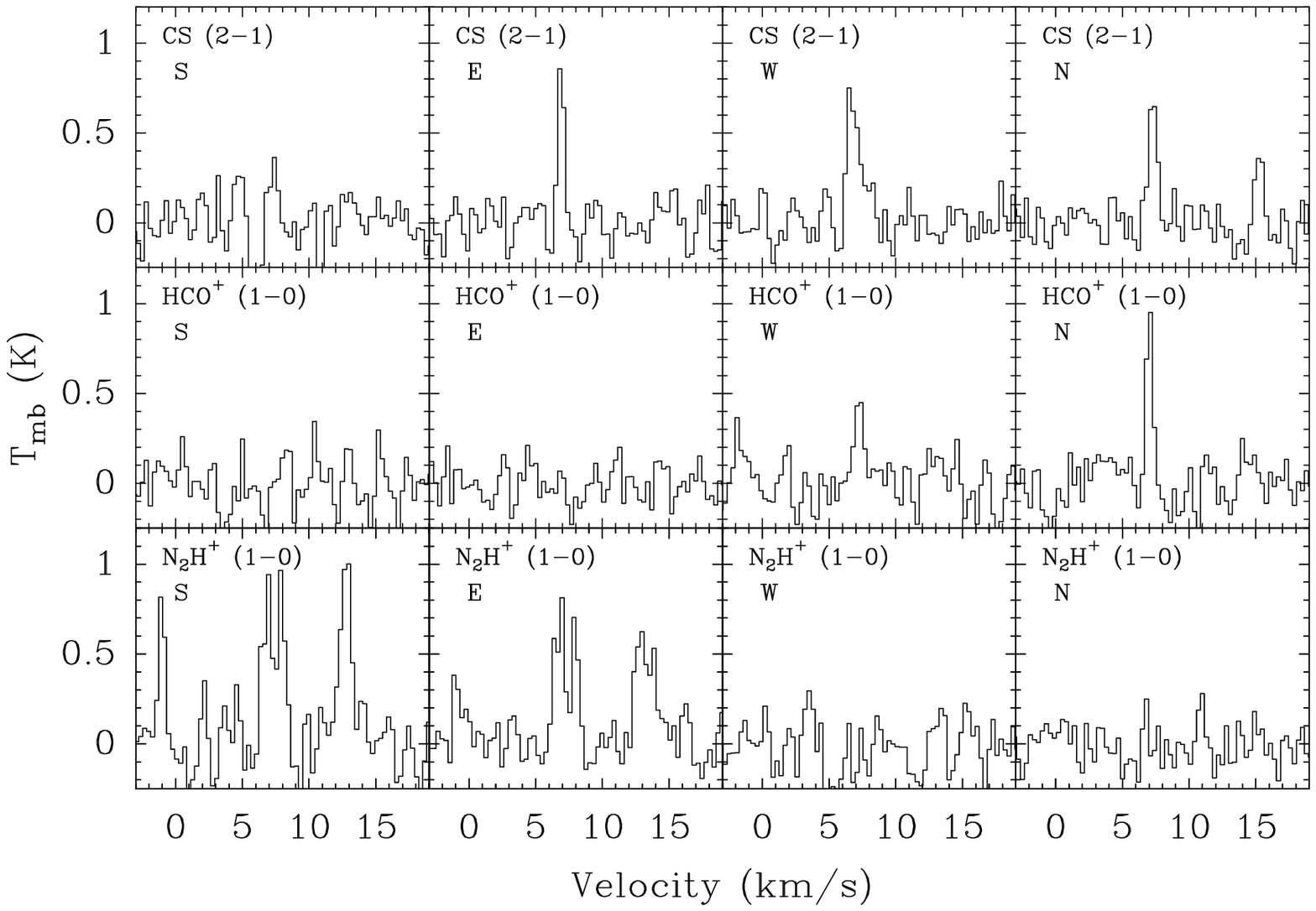}
\end{center}
\caption{Spectra of the CS (\jr{2}{1}), HCO$^+$ \mbox{($J$=1$\rightarrow$0)},
and N$_2$H$^+$ \mbox{($J$=1$\rightarrow$0)} transitions at the four selected
positions in our mapped region.}
\label{fig:spectra20}
\end{figure}

\begin{table*}[t]
\caption{CS (\jr{2}{1}), HCO$^+$ \mbox{($J$=1$\rightarrow$0)}, and N$_2$H$^+$
\mbox{($J$=1$\rightarrow$0)} lines and physical parameters obtained at the
four selected positions in our region}
\label{tab:parameters}
\begin{center}
\includegraphics[width=15cm]{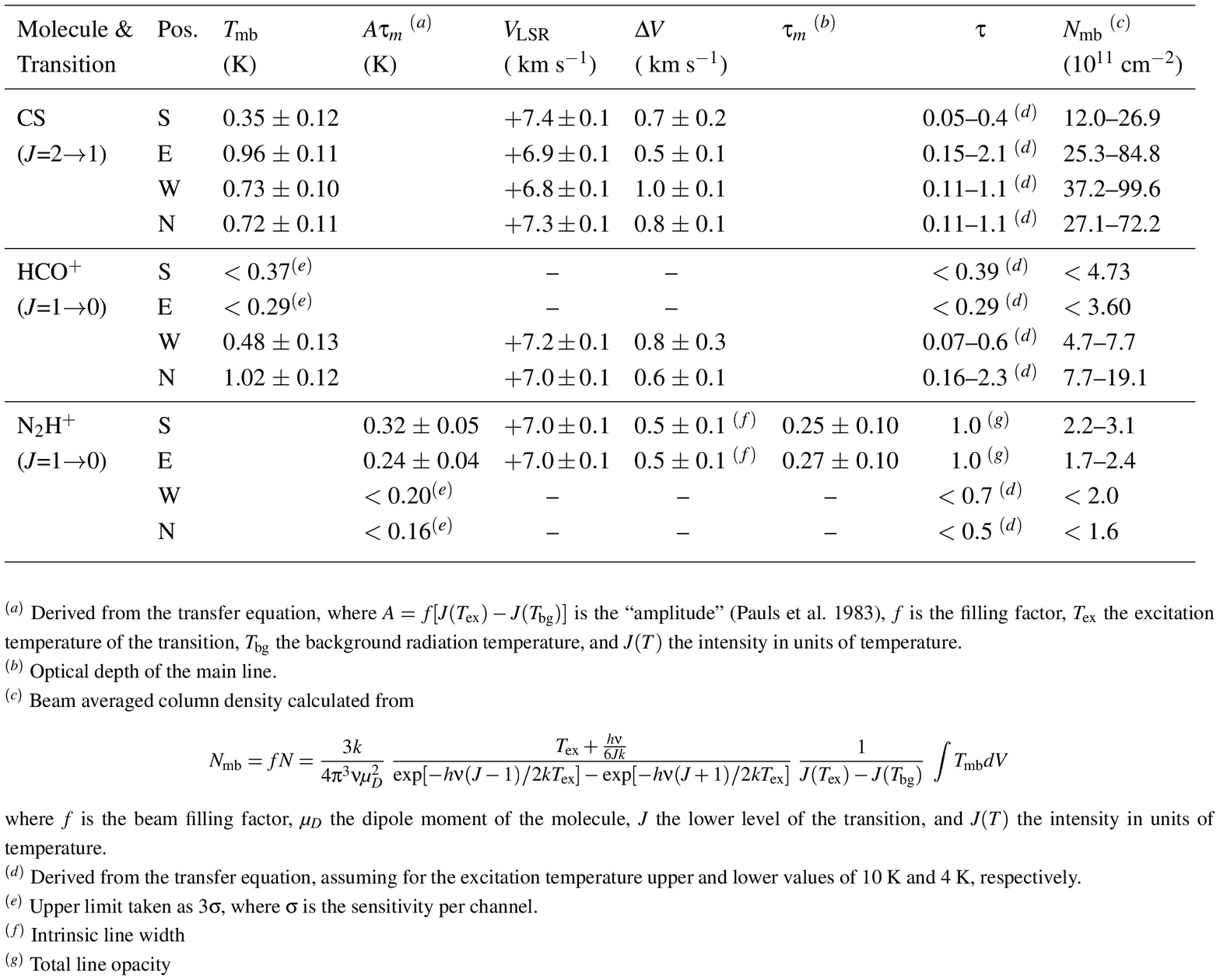}
\end{center}
\end{table*}

Figure~\ref{fig:spectra20} shows the spectra obtained for the three
transitions detected in the region in the four selected
points. Table~\ref{tab:parameters} lists the fitted line parameters for each
detected transition. CS emission is detected, with varying intensity, in all
four positions, whereas HCO$^+$ emission is only detected at the N and W
positions, and N$_2$H$^+$ emission at the S and E positions. Line center
velocities for CS and HCO$^+$ at the positions where a reliable fit could be
obtained are compatible with being originated in the same bulk of gas. The
difference in the line center velocity is $\sim0.3$--0.5~km~s$^{-1}$, which is
the velocity resolution of our observations. However, line center velocities
for the N$_2$H$^+$ spectra were shifted $>0.7$~km~s$^{-1}$ from the CS line
velocities. Using the much better determined frequency for the main N$_2$H$^+$
\mbox{($J$=1$\rightarrow$0)} component ($F_1,F=2,3\rightarrow1,2$) given by
Caselli et al.  (\cite{caselli}): 93.17378 GHz, we found a good agreement with
CS spectra line velocities, $\la 0.3 $~km~s$^{-1}$. We have corrected the
N$_2$H$^+$ line velocities accordingly, as Table~\ref{tab:parameters}
shows. Line-widths are $< 0.8$~km~s$^{-1}$, very similar to the
observed values in the single-dish observations, except for the W clump where
velocity dispersion is higher, $\sim 1.0$~km~s$^{-1}$.

\subsection{Physical parameters}

We were able to estimate the excitation temperature for the N$_2$H$^+$ data at
the two positions, S and E, where we had good signal-to-noise spectra. The
values obtained at both positions are $\sim 4$~K (assuming a beam filling
factor $f=1$), which is compatible with the value obtained from the CS
\mbox{($J$=1$\rightarrow$0)} observations, 4.2~K (Morata et
al. \cite{morata}), and slightly lower than the value obtained from the NH$_3$
\mbox{($J,K$)=(1,1)} observations, 5.7~K (Sep\'ulveda et al.
\cite{sepulveda}). The total line opacity of the N$_2$H$^+$
\mbox{($J$=1$\rightarrow$0)} transition at both positions is $\sim 1.0 $. We
estimated an upper limit for the beam averaged column density using the value
for the excitation temperature obtained from the CS
\mbox{($J$=1$\rightarrow$0)} and N$_2$H$^+$ \mbox{($J$=1$\rightarrow$0)}
observations, $T_{\rm ex}=4$~K, and assuming a filling-factor $f=1$.  However,
the filling-factor could be $f<1$. In this case, the excitation temperature
would be greater than 4~K, but likely less than 10~K, because the kinetic
temperature in low-mass dense cores is typically about 10~K, and the
excitation temperature will not be greater than this value. Thus, we used an
upper value for the excitation temperature $T_{\rm ex}=10$~K, which implies a
lower limit for the beam average column density. The values for the line
opacity, $\tau$, and the beam averaged column density, $N_{\rm mb}$ for the
four selected positions are shown in Table~\ref{tab:parameters}.

\subsection{Comparison with single-dish observations}

\begin{figure}[t]
\begin{center}
\includegraphics[width=\hsize]{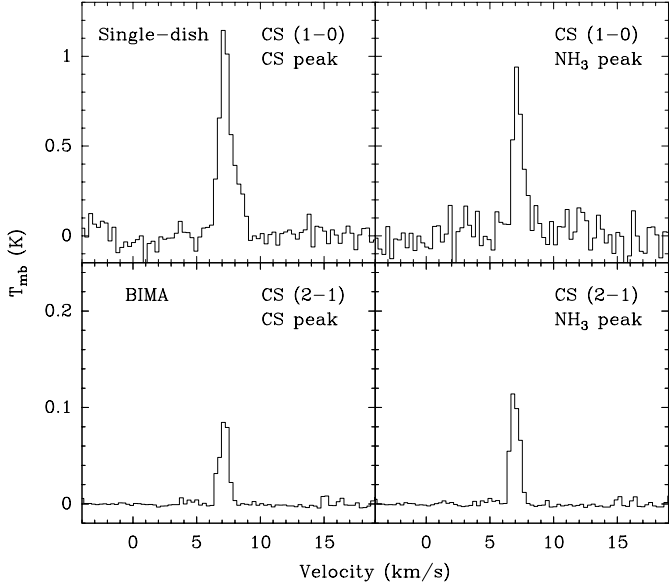}
\end{center}
\caption{Spectra of the CS \mbox{($J$=1$\rightarrow$0)} transitions obtained
at the positions of the CS \mbox{($J$=1$\rightarrow$0)} and NH$_3$
\mbox{($J,K$)=(1,1)} single-dish emission peaks with the Yebes telescope
(top), compared with the spectra of the CS \mbox{($J$=2$\rightarrow$1)}
transition obtained at the same positions after convolving the BIMA data with
a beam of $1\farcm9$ equal to the single-dish beam (bottom). Note the
different vertical scales.}
\label{fig:spectrayeb}
\end{figure}

We compared the CS (\jr{2}{1}) interferometric and
\mbox{($J$=1$\rightarrow$0)} single-dish
emissions. Figure~\ref{fig:spectrayeb} shows the single-dish CS
\mbox{($J$=1$\rightarrow$0)} spectra obtained at the positions of the CS
\mbox{($J$=1$\rightarrow$0)} and NH$_3$ \mbox{($J,K$)=(1,1)} emission peaks,
compared with the BIMA CS (\jr{2}{1}) spectra obtained at the same positions
after convolving the data with a gaussian function to obtain a $1\farcm9$
resulting beam, equal to the beamsize of the single-dish observations. Line
center velocities agree within 0.3~km~s$^{-1}$, our spectral resolution in
both sets of observations. Line widths are almost coincident at the southern
position, while the \mbox{($J$=1$\rightarrow$0)} line is $\sim
0.35$~km~s$^{-1}$ wider at the northern position. In order to estimate the
flux loss in our interferometric observations, we calculated the line
intensity we should obtain for the \mbox{($J$=2$\rightarrow$1)} line from the
\mbox{($J$=1$\rightarrow$0)} line opacity and excitation temperature obtained
previously in the single-dish observations (Morata et al. \cite{morata}). We
found that the measured line intensity is $\sim8$\% in the northern position,
and $\sim12$\% in the south position from the estimated values. These results
are compatible with the characteristics of the BIMA array, which is
insensitive to structures larger than about 10 times the synthesized beamwidth
(Wright \cite{wright}), which in our case is $\sim1\farcm5$--$2'$ (0.13--0.17
pc), of the order of the single-dish beamsize.

However, the existence of such an extended component, responsible for
$\sim90\%$ of the single-dish line intensity,  would not affect significantly 
the beam averaged column densities determined from our interferometric 
observations. Assuming a size of $120''$ for the extended component, its
contribution to the beam averaged column density for a beam size of $17''$
would be of the order of $90\%\times(17''/120'')^2=2\%$.  Thus, the beam
averaged column densities determined from our observations are not
significantly affected by the missing emission by the BIMA interferometer,
assuming that it is due to an extended cloud component.

\begin{figure}[t]
\begin{center}
\includegraphics[angle=270,width=\hsize]{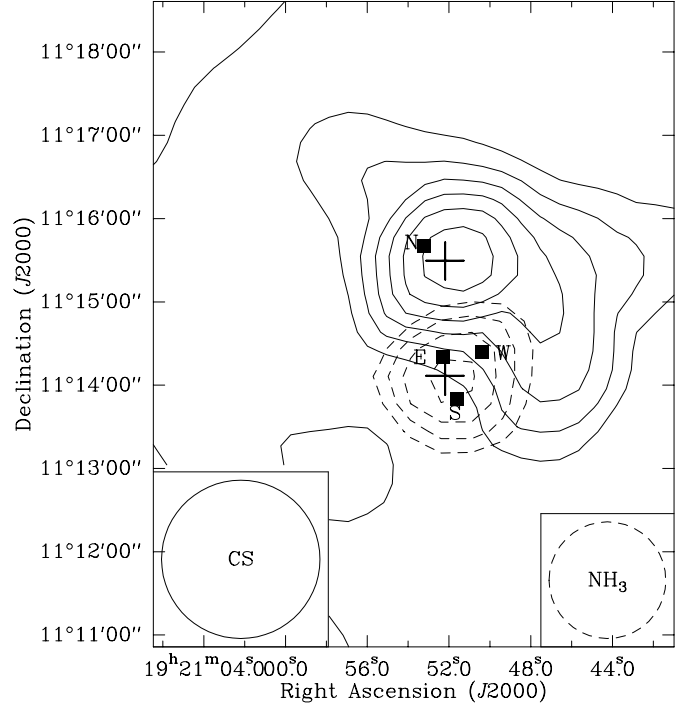}
\end{center}
\caption{Contour map of the single-dish integrated line intensity of the CS
\mbox{($J$=1$\rightarrow$0)} line (solid line) obtained with the Yebes
telescope (Morata et al.  \cite{morata}), and the single-dish main beam
brightness temperature of the NH$_3$ \mbox{($J,K$)=(1,1)} transition (dashed
line) obtained with the Haystack telescope (Sep\'ulveda et
al. \cite{sepulveda}). The lowest CS contour level is 0.48~K~km~s$^{-1}$, and
the increment is 0.05~K~km~s$^{-1}$. The lowest NH$_3$ contour level is 1.35
K, and the increment is 0.3~K. The crosses indicate the nominal position of
the single-dish emission peaks of the CS \mbox{($J$=1$\rightarrow$0)} line
(north) and NH$_3$ \mbox{($J,K$)=(1,1)} line (south). The filled squares
indicate the position of the four points selected for further study in the
interferometric observations}
\label{singledish}
\end{figure}

Figure~\ref{singledish} shows the distribution of the single-dish CS
\mbox{($J$=1$\rightarrow$0)}~(Morata et al. \cite{morata}) and NH$_3$
\mbox{($J,K$)=(1,1)} (Sep\'ulveda et al. \cite{sepulveda}) emission in the
same region mapped by the interferometric observations. We found that most of
the strongest CS (\jr{2}{1}) emission and particularly all the detected
N$_2$H$^+$ emission were enclosed inside the half-maximum contours of the
NH$_3$ map, located at the southern region of the map. All four relative
emission enhancements in these two transitions are also enclosed inside the
beam of the single-dish observations. HCO$^+$ emission was fainter around this
southern region, and marginal in the eastern, western and southern margins of
the NH$_3$ intensity contours. Around the northern region, no N$_2$H$^+$
emission was detected, but CS and HCO$^+$ emission were enclosed by the beam
of the CS single-dish observations. Emission around this position was not as
concentrated as around the southern region, but we found intense HCO$^+$ and
CS (\jr{2}{1}) emission very near the beam center.

\section{Discussion}

\subsection{Structure of the core}

The integrated intensity maps of the CS (\jr{2}{1}),
N$_2$H$^+$~\mbox{($J$=1$\rightarrow$0)}, and HCO$^+$
\mbox{($J$=1$\rightarrow$0)} transitions obtained with the BIMA interferometer
showed that a much clumpier medium is revealed by the high angular resolution
observations than by the single-dish maps of CS \mbox{($J$=1$\rightarrow$0)}
(Morata et al. \cite{morata}) and NH$_3$ \mbox{($J,K$)=(1,1)} (Sep\'ulveda et
al. \cite{sepulveda}) obtained with an angular resolution of $\sim
1\farcm5$. Several clumps of size $\la0.08$ pc are found distributed all over
the region where the strongest emission of the single-dish observations was
located, which would support the idea that at least part of the single-dish
emission was originated in clumps of smaller size, $<0.1$ pc (Taylor et
al. \cite{tayloru}). Moreover, these small size clumps, although coinciding in
some cases, are traced differently by each molecule, which would support the
idea of a chemical differentiation between each clump of gas depending on the
local density and age. However, we found that $\sim90\%$ of the single-dish CS
emission is not detected by the interferometric observations. As we have
already pointed out, the BIMA array is insensitive to structures larger than
$\sim0.17$ pc, which could mean that part of the emission could come from
chemically young clumps of that size or larger, undetectable then by BIMA, but
whose emission was detected by the single-dish observations. Another
possibility is that there could also be clumps smaller than the ones we found,
of low density, and with emission not strong enough to be detected with the
interferometer.

In order to qualitatively test which of the aforementioned two possibilities
is more consistent with our observations, we modeled the filtering effect of
the BIMA interferometer response to a molecular core with an homogeneous and
extended geometry and a clumpy heterogeneous geometry. The process of
generating synthetic visibility data matching that expected from true
observations with the BIMA interferometer was performed by using the MIRIAD
task UVGEN. This process is described by Girart \& Acord
(\cite{giracord}). Our synthetic observations had the same phase center as the
L673 observations and were done for the BIMA C-array configuration. The total
flux adopted for both cases, the extended and clumpy structures, was similar
to that expected for the CS ($J$=2$\to$1) transition, calculated from the
measured CS ($J$=1$\to$0) intensity. The results obtained were:
\begin{enumerate}

\item {\it Extended and homogeneous medium.}
The presence of the extended emission was modeled as single large component. We
tested the filtering effect of the interferometer for different sizes and
different center positions of the cloud with respect to the phase center. We
found that in order to achieve a missing flux of about 90\%, the value
estimated for the BIMA observations, large sizes were required: diameter 
$\ga2'$. The resulting synthetic maps showed strong negative lobes extending
$\sim40''$ along the N-S direction, at both sides of the center of the
synthetic cloud. This was due to a combination of the effects of source
structure, with a large fraction of undetected flux by the interferometer, and
visibility coverage of the BIMA interferometer for a source of low declination
($\delta=11\degr$ for L673). These negative structures could not be removed
in the cleaning process. The cleaned maps had absolute peak intensities of the
negative structures with similar, or even higher, values than the positive
structures. 

However the real BIMA CS ($J$=2$\to$1) channel maps does not show such
strong negative structures. At the central velocities of the CS line there
are some negative lobes, but their intensity is only a $\la40\%$ of the
positive peaks.

\item {\it Clumpy and heterogeneous medium.}
The chemical models of Taylor et al.\ (\cite{tayloru}; \cite{taylord}) assume
that the clouds are formed by clumps of $\la0.1$~pc with different sizes,
ages and masses. From this assumption it is expected a cloud formed with only
few massive clumps and several clumps less massive. So, we adopted a
simplistic model of a cloud formed with one massive, thereby strong, clump and
between 15 and 20 less massive, thereby weaker, clumps. We found that even
most of the emitting flux came from the smaller clumps, the BIMA synthetic
resulting maps were clearly dominated by the strongest clump, and a large
fraction of the emission from the weaker clumps was undetected by the
interferometer. In some of the cases modeled, the total missing flux was of
$\sim80\%$ of the total flux. The residual negative structures of the
synthetic maps were 30--50\% of the positive.

\end{enumerate}

In summary, without going to an exhaustive analysis of the response of the BIMA
interferometer to the different cloud structures, we find that in L673 the
missing flux is most probably due to a clumpy heterogeneous structure.

\subsection{Chemical abundances}

In our maps, we found that all the detected N$_2$H$^+$ emission was enclosed
inside the half-maximum contours of the NH$_3$ map of Sep\'ulveda et al.
(\cite{sepulveda}), with the strongest N$_2$H$^+$ emission found inside the
highest NH$_3$ contours. The nominal position of the NH$_3$ emission peak is
placed between the two local N$_2$H$^+$ emission peaks, although the beam
enclosed both maxima. The chemical model developed by Taylor et
al. (\cite{tayloru}; \cite{taylord}) marks the NH$_3$ and N$_2$H$^+$ molecules
as forming at late times, when sufficiently high densities are reached or in
long lived clumps. Thus, they should be found in the same region. The
coincidence between the distribution of the emission of N$_2$H$^+$ observed by
the BIMA interferometer and the NH$_3$ emission detected by single-dish
observations, and the overall morphological appearance of the NH$_3$ and
N$_2$H$^+$ emission seem to support these predictions. Moreover, both
interferometric and single-dish emissions are highly concentrated, and are
compatible with being originated in the same gas since the emission of
N$_2$H$^+$ could be coming from two clumps of size $\sim 0\farcm6 \times 1'$
enclosed inside the higher levels of the lower resolution emission of
NH$_3$. The SO molecule was also marked in our model as a late-time
molecule. The near coincidence of the clump found in the marginally detected
SO emission with the northern clump of the N$_2$H$^+$ emission would support
that classification.

\begin{table}[t]
\caption{Column density ratios}
\label{tab:ratios}
\begin{center}
{\small
\begin{tabular}{lccc}
\noalign{\bigskip}\hline\noalign{\smallskip}
Position & [HCO$^+$/CS] & [N$_2$H$^+$/CS] & N(CS)/N$_{\rm max}$(CS)\\
\noalign{\smallskip}\hline\noalign{\medskip}
S & $<0.21$ & ~~~0.15 & 30\%\\
E & $<0.07$ & ~~~0.05 & 77\%\\
W & ~~~0.10 & $<0.02$ & 100\%\,\,\,\\
N & ~~~0.27 & $<0.02$ & 73\%\\
\noalign{\medskip}\hline\noalign{\bigskip}
\end{tabular}
}
\end{center}
\end{table}

Table~\ref{tab:ratios} shows the column density ratios of N$_2$H$^+$ and
HCO$^+$ over CS for the four selected positions calculated from the column
density determinations of Table~\ref{tab:parameters}, and the CS column
density relative to the maximum value, found in position W. We observe that
N$_2$H$^+$ abundance is $\sim$ 5--15 \% of that of CS for the S and E
positions, and becomes lower at the N and W positions. [HCO$^+$/CS] is $\la 20
$\% for the S, E, and W positions, and becomes $\sim {\bf 30 } $\% at the N
position, while there is a factor 3 variation in the CS column density. We can
then differentiate between the N and W positions, where HCO$^+$ is clearly
detected while N$_2$H$^+$ is not, and the E and S positions where N$_2$H$^+$
is clearly detected whereas HCO$^+$ is not detected.

\subsection{Modeling the chemical evolution}

\begin{figure}[t]
\begin{center}
\includegraphics[width=\hsize]{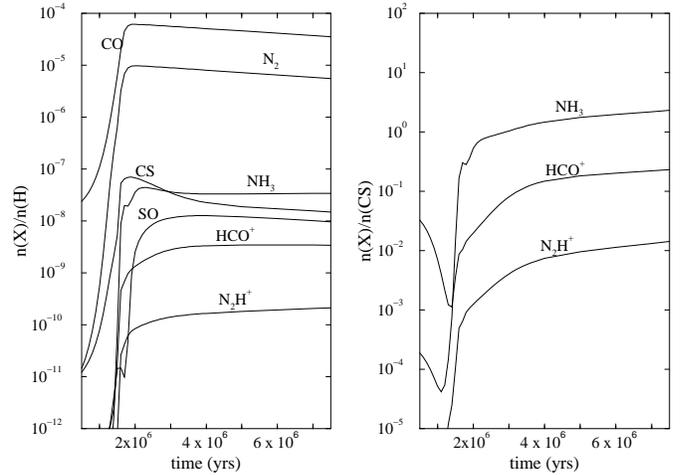}
\end{center}
\caption{{\it Left panel:} Chemical fractional abundances (relative to H
nuclei) as a function of time for a free-fall collapse model halted at density
$n_{\rm H}=5\times10^4$~cm$^{-3}$. Initially $n_{\rm H}=1\times10^3$~cm$^{-3}$
and $A_V=0.5$. Freeze-out parameter $FR=0.01$, corresponding to an average
value of the product of the dust to gas number density ratio and square of the
grain radius $\langle n_d a^2\rangle=2\times 10^{-20}$~cm$^{-2}$. {\it Right
panel:} Fractional abundances of selected species (relative to CS) as a
function of time.}
\label{sdtmodel}
\end{figure}

We used the results of the model explained in Taylor et al. (\cite{taylord})
to compare the predicted abundances of HCO$^+$ and N$_2$H$^+$ with respect to
CS, with the values we obtained for the column density ratios, and see if some
chemical age determination could be made for these
positions. Figure~\ref{sdtmodel} shows the predicted fractional abundances for
several molecules of interest for values of collapse, $B=1$, freeze-out,
$FR=0.01$, and final density $n_f=5\times10^4$~cm$^{-3}$ (see Taylor et
al. \cite{tayloru}; \cite{taylord}). We also show the molecular abundance
relative to CS for the HCO$^+$, N$_2$H$^+$, and NH$_3$ molecules.

We found that the abundance [HCO$^+$/CS] can be explained by the model in
relatively short times, between 2-3$\times10^6$ years. Thus, a higher HCO$^+$
fractional abundance would indicate a more chemically evolved clump. However,
this does not seem to agree with the abundance of N$_2$H$^+$. First,
N$_2$H$^+$ reaches high abundances at later times, and one would expect to
find N$_2$H$^+$ at more evolved clumps. Second, the model can not explain
fractional abundances of N$_2$H$^+$ similar to those of HCO$^+$, and as high
as those we found in the S and E positions. A slightly higher freeze-out
parameter would provide higher relative N$_2$H$^+$ abundance with respect to
CS, while not changing appreciably the fractional abundance, because CS is
frozen-out to grains faster. In this case, we also obtain higher HCO$^+$
relative abundances with CS.

\begin{figure}[t]
\begin{center}
\includegraphics[width=\hsize]{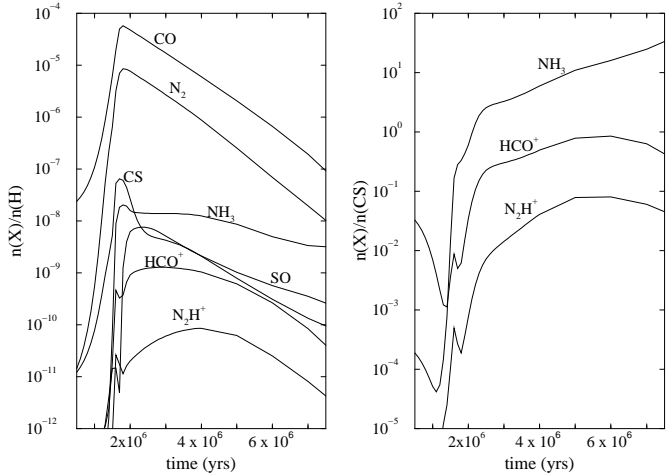}
\end{center}
\caption{{\it Left panel:} Chemical fractional abundances (relative to H
nuclei) as a function of time for a free-fall collapse model halted at density
$n_{\rm H}=5\times10^5$~cm$^{-3}$. Initially $n_{\rm H}=1\times10^3$~cm$^{-3}$
and $A_V=0.5$. Freeze-out parameter $FR=0.01$, corresponding to an average
value of the product of the dust to gas number density ratio and square of the
grain radius $\langle n_d a^2 \rangle =2\times 10^{-20}$~cm$^{-2}$. {\it Right panel:}
Fractional abundances of selected species (relative to CS) as a function of
time.}
\label{noumodel}
\end{figure}

Figure~\ref{noumodel} shows the results of the model modifying the final
density value to 10 times the previous value, $n_f=5\times10^5$~cm$^{-3}$, and
maintaining unchanged the values of $B$ and $FR$. We find some differences
with the previous results. CS maximum fractional abundance is similar to the
previous model, while NH$_3$, HCO$^+$, and N$_2$H$^+$ reach slightly lower
abundances. In any case, CS and NH$_3$ show the same behavior found in the
model of Taylor et al. (\cite{tayloru}). However, the higher final density
produces a faster depletion of molecules, especially CS which has a steeper
decline with time.

At times $\sim2.5\times10^6$ years, when CS has a high fractional abundance,
and HCO$^+$ has barely reached its maximum fractional abundance, we find a
relative abundance [HCO$^+$/CS]$\sim0.1$, while N$_2$H$^+$ has still a low
abundance. This could correspond to the W position of our map. At a later
time, $\sim3\times10^6$ years, when N$_2$H$^+$ is still at low abundances,
HCO$^+$ abundance is practically the same while CS abundance is slightly
lower, with a resulting higher [HCO$^+$/CS] ratio.  This could correspond to
the N position. At the time N$_2$H$^+$ reaches its peak abundance,
$\sim4\times10^6$ years, HCO$^+$ has diminished in abundance although it can
still be observed. This would point to the S and E positions being at later
stages of chemical evolution. Distinguishing between the two is difficult. The
S position could be more evolved since it shows a lower CS column density, and
a higher [N$_2$H$^+$/CS] ratio, which could correspond to the stage where
N$_2$H$^+$ is at higher abundances. 

However, even the modified model is not able to explain the observed HCO$^+$
low abundances obtained at positions S and E with respect to the predicted
values. The HCO$^+$ molecule has a high dipole moment, and it has been found
that the \mbox{($J$=1$\rightarrow$0)} transition line profiles are often
self-absorbed in typical molecular cloud cores by foreground gas at lower
density and temperature. The apparently sudden lack of HCO$^+$ emission in
those places where N$_2$H$^+$ or CS emission is intense may suggest that the
HCO$^+$ \mbox{($J$=1$\rightarrow$0)} emission arising from the core might be
absorbed efficiently by a cold low-density envelope around the core or a
foreground cloud, as in the case of the NGC 2264G region (Girart et
al. \cite{girart}).

\subsection{Comparison with other studies}

Studies of chemical evolution have been made up to now with single-dish
observations using several transitions of selected early- and late-time
molecules, usually CS, N$_2$H$^+$, NH$_3$, HCO$^+$, SO, SO$_2$, and HC$_3$N.
These studies of starless cores show maps where emission seems to originate in
several clumps distributed all over the region (Dickens et al.\
\cite{dickens}) or in round-like clumps (Tafalla et al.\ \cite{tafalla}). All
these studies show evidences of chemical differentiation. Dickens et al.\ find
that early- and late-time molecules are found at different positions and show
differing behavior, i.e. early-time molecules are most abundant where
late-time molecules are not, and viceversa.  The authors explain their results
invoking a difference in the chemical ages of different parts of their region.
Tafalla et al.\ also find that their maps show a great coincidence between the
emission of the N$_2$H$^+$ and NH$_3$ molecules, while differing from the CS
emission.

Some of these studies also find evidence for differing depletion
rates of molecules such as CS or NH$_3$. Tafalla et al.\ find a central
abundance drop of CS, a constant abundance of N$_2$H$^+$, and a central
enhancement of NH$_3$ in their regions, while Bergin et al.\ (\cite{bergin}),
in observations of the IC 5146 cloud, also detect molecular depletion of CS,
while N$_2$H$^+$ remains in the gas phase with growing extinction. Other
numerical calculations, such as Aikawa et al.\ (\cite{aikawa}) also predict a
depletion of certain species from the central regions. In our model, we would
expect to find the more massive and chemically older clumps in the central
regions. These clumps would have substantial quantities of NH$_3$ and
N$_2$H$^+$, while CS would have been depleted from the gas phase, which would
show as a central depletion of CS.

Thus, these results suggest that although our interferometric observations lose
information on the total emission of the region,  they are in agreement with
the results obtained with single-dish telescopes, suggesting that
interferometric observations are useful to study with high angular resolution
the chemical evolution of inner structures of starless cores.

\section{Summary}

We made a multitransitional study with the BIMA interferometric array of the
starless core found in the L673 (Morata et al. \cite{morata}), in order to
test the chemical model of Taylor et al. (\cite{tayloru}; \cite{taylord}). The
main results were:

\begin{enumerate}

\item We detected emission in the CS(\jr{2}{1}), N$_2$H$^+$
\mbox{($J$=1$\rightarrow$0)}, and HCO$^+$ \mbox{($J$=1$\rightarrow$0)}
lines. We marginally detected emission in the SO ($J_K=2_3\rightarrow1_2$)
line.

\item The high angular resolution interferometric observations revealed a much
clumpier medium than the lower resolution single-dish observations. Several
clumps of size $\la0.08$ pc were found for each line distributed all over the
region where the single-dish emission was found.

\item Each molecular transition traced differently the clump distribution,
although in some cases the detected clumps were coincident. We found that the
distribution of the N$_2$H$^+$ emission was completely enclosed by the
half-maximum contours of the NH$_3$ maps of Sep\'ulveda et al.
(\cite{sepulveda}), and that both emissions were highly compact and compatible
with being originated in the same gas.
We also found a marginal coincidence with the SO molecule, which is also a
late-time molecule.

\item The BIMA interferometer detected only 9--12\% of the CS
(2$\rightarrow$1) emission. Modelling the filtering effect of BIMA, we found
that a clumpy and heterogeneous medium could explain this effect.

\item We estimated the abundance ratios [N$_2$H$^+$/CS] and [HCO$^+$/CS] from
the molecular column density determinations in four selected positions,
labeled as S, E, W, and N, in order to compare them with the predicted values
of the chemical model. We found that at the N and W positions there was a high
HCO$^+$ abundance relative to CS and a low N$_2$H$^+$ abundance, while at the
S and E positions the case was reversed.

\item The chemical model was able to explain the abundance ratios of
[HCO$^+$/CS] at the positions where HCO$^+$ was detected, N and W. The best
fit was for a model with a density at which collapse is halted of
$n_f=5\times10^5$~cm$^{-3}$. In this case, we found that the predicted
[N$_2$H$^+$/CS] abundance ratios were more in agreement with the
observations. A high HCO$^+$ abundance, but a low N$_2$H$^+$ abundance would
represent an earlier stage, where the W and N positions could be found. The S
and E positions would be in a later chemical stage on account of the higher
N$_2$H$^+$ abundance, although a more precise differentiation is hard to make.

\end{enumerate}

Thus, the model described in Taylor et al. (\cite{tayloru}; \cite{taylord})
applied to a starless core with no apparent signs of other star formation
tracers, such as molecular outflows or radio continuum sources, was able to
explain the interferometric observations, if we take into account the likely
self-absorption of HCO$^+$. In particular, the chemical differentiation of the
detected clumps and the coincidence of the NH$_3$ and N$_2$H$^+$ emissions
agree with the model. It also enabled us to explain the HCO$^+$ and N$_2$H$^+$
emissions and to propose a classification of the studied clumps according to
the stage of chemical evolution indicated by the molecular abundances.

Further multitransitional observations with higher sensitivity would be
necessary in order to determine with more detail the physical structure of the
region and to provide more data with which to refine the chemical model
predictions. In particular, the observation of molecules such as H$_2$CO and
SO, or of the isotope H$^{13}$CO$^+$, would help to clarify the nature
of the HCO$^+$ \mbox{($J$=1$\rightarrow$0)} emission.

\begin{acknowledgements}
JMG acknowledges support by NSF grant AST-99-81363 and by RED-2000 from the
Generalitat de Catalunya.  RE and JMG are partially supported by DGICYT grant
PB98-0670 (Spain).
We thank the referee for the comments and discussion.
\end{acknowledgements}


\begin{thebibliography}{}

\bibitem[2001]{aikawa}Aikawa, Y., Ohashi, N., Inutsuka, S.-I., Herbst, E., \&
Takakuwa, S. 2001, ApJ, 552, 639

\bibitem[1986]{beichman}Beichman, C.~A., Myers, P.~C., Emerson, J.~P., et al.
1986, ApJ, 307, 337

\bibitem[1989]{benson}Benson, P.~J., \& Myers, P.~C. 1989, ApJSS, 71, 89

\bibitem[2001]{bergin}Bergin, E.~A., Ciardi, D.~R., Lada, C.~J., Alves, J., \&
Lada, E.~A. 2001, ApJ, 557 209

\bibitem[1995]{caselli}Caselli, R., Myers, P.~C., \& Thaddeus, P. 1995, ApJ,
455, L77

\bibitem[2000]{dickens}Dickens, J.~E., Irvine, W.~M., Snell, R.~L., et
al. 2000, ApJ, 542, 870

\bibitem[2001]{giracord} Girart, J.~M. \& Acord, J.~M. P. 2001, ApJ, 552, L63

\bibitem[2000]{girart}Girart, J.~M., Estalella, R., Ho, P.~T.~P., \& Rudolph,
A.~L. 2000, ApJ, 539, 763

\bibitem[1983]{herbig} Herbig, G.~H., \& Jones, B.~F. 1983, AJ, 88, 1040

\bibitem[1997]{morata}Morata, O., Estalella, R., L\'opez, R., \& Planesas,
P. 1997, MNRAS, 292, 120

\bibitem[1991]{myers}Myers, P.~C., Fuller, G.~A., Goodman, A.~A., \& Benson,
P.~J. 1991, ApJ, 376, 561

\bibitem[1991]{pastor}Pastor, J., Estalella, R., L\'opez, R., et al. 1991, A\&A,
252, 320

\bibitem[1983]{pauls}Pauls, T.~A., Wilson, T.~L., Bieging, J.~H., \& Martin,
R.~N. 1983, A\&A, 123, 23

\bibitem[1995]{sault}Sault, R.~J., Teuben, P.~J., \& Wright, M.~C.~H. 1995, in
ASP Conf.\ Ser. 77, Astronomical Data Analysis Software and Systems IV,
ed. R.~A. Shaw, H.~E. Payne, \& J.~J.Hayes, San Francisco: Astronomical
Society of the Pacific, 433

\bibitem[2002]{sepulveda}Sep\'ulveda, I., et al. 2002, in preparation

\bibitem[2002]{tafalla}Tafalla, M., Myers, P.~C., Caselli, P., Walmsley,
C.~M., \& Comito, C. 2002,  ApJ, 569, 815

\bibitem[1996]{tayloru}Taylor, S.~D., Morata, O., \& Williams, D.~A. 1996, A\&A, 313, 269

\bibitem[1998]{taylord}Taylor, S.~D., Morata, O., \& Williams, D.~A. 1998, A\&A, 336, 309

\bibitem[1996]{wright}Wright, M.~C.~H. 1996, BIMA memo 45

\bibitem[1989]{zhou}Zhou, S., Wu, Y., Evans, N.~ J., Fuller, G.~ A., \& Myers,
P. C. 1989, ApJ, 346, 168

\end{thebibliography}
\end{document}